\documentclass[11pt]{article}
\usepackage{alltt,verbatim,vmargin,graphicx,amsmath,calc,rotating}
\usepackage{subfigure}
\usepackage{epsfig}

\input pubboard/babarsym

\newcommand {\brhoenu}{$\ensuremath{B^+ \rightarrow \rho^0 e^+ \nu}$}
\newcommand {\brhochgenu}{$\ensuremath{B^0 \rightarrow \rho^- e^+ \nu}$}
\newcommand {\bpienu}{$\ensuremath{B^+ \rightarrow \pi^0 e^+ \nu}$}
\newcommand {\bpichgenu}{$\ensuremath{B^0 \rightarrow \pi^- e^+ \nu}$}
\newcommand {\bomegaenu}{$\ensuremath{B^+ \rightarrow \omega e^+ \nu}$}

\newcommand{\pl}{{\sl Phys. Lett.~}}
\newcommand{\prl}{{\sl Phys. Rev. Lett.~}}

\setpapersize{USletter}
\setmarginsrb{1.1in}{0.5in}{1.1in}{0.5in}%
{1.0\baselineskip}{1.5\baselineskip}{1.5\baselineskip}{2.5\baselineskip}

\long\def\inst#1{\par\nobreak\kern 4pt\nobreak
    {\it #1}\par\vskip 10pt plus 3pt minus 3pt}

\begin{document}
{\pagestyle{empty}

\begin{flushright}
\babar-CONF-02/030\\
SLAC-PUB-9305\\
July 2002
\end{flushright} 
\bigskip \bigskip \bigskip 
\bigskip

\begin{center} \Large \bf
Measurement of the CKM Matrix Element {\boldmath $|V_{ub}|$} \\
with Charmless Exclusive Semileptonic {\boldmath $B$} Meson Decays\\
at \babar\ 
\end{center} \bigskip

\begin{center}
\large The \babar\ Collaboration\\
\mbox{ }\\
July 25, 2002
\end{center} 
\bigskip
\bigskip

\begin{center}
\large \bf Abstract
\end{center}

  We present a preliminary measurement of the branching fraction for
  $B \rightarrow \rho e \nu$ and of the CKM matrix element $|V_{ub}|$
  using approximately $55$ million $B\bar{B}$ meson pairs collected
  with the \babar\ detector. Using isospin relations for several
  modes we find
  \begin{eqnarray}
  {\cal B}(B^0\rightarrow~\rho^-~e^+~\nu) & = & 
  (3.39~\pm~0.44~\pm~{0.52}\pm~0.60)~\times 10^{-4} \nonumber  \\
  |V_{ub}| & = & (3.69~\pm~0.23~\pm~0.27~{}^{+0.40}_{-0.59})~\times~10^{-3}. \nonumber
  \end{eqnarray}
  The quoted errors are statistical, systematic, and theoretical 
  respectively. These results are obtained by using five different 
  form-factor calculations.

\vfill
\begin{center}
Contributed to the 31$^{st}$ International Conference on High Energy Physics,\\ 
7/24---7/31/2002, Amsterdam, The Netherlands
\end{center}

\vspace{1.0cm}
\begin{center}
{\em Stanford Linear Accelerator Center, Stanford University, 
Stanford, CA 94309} \\ \vspace{0.1cm}\hrule\vspace{0.1cm}
Work supported in part by Department of Energy contract DE-AC03-76SF00515.
\end{center}

\newpage
} 

\begin{center}
\small

The \babar\ Collaboration,
\bigskip

B.~Aubert,
D.~Boutigny,
J.-M.~Gaillard,
A.~Hicheur,
Y.~Karyotakis,
J.~P.~Lees,
P.~Robbe,
V.~Tisserand,
A.~Zghiche
\inst{Laboratoire de Physique des Particules, F-74941 Annecy-le-Vieux, France }
A.~Palano,
A.~Pompili
\inst{Universit\`a di Bari, Dipartimento di Fisica and INFN, I-70126 Bari, Italy }
% ...

\end{center}\newpage

\section{Introduction}

Exclusive $b \rightarrow u \ell \nu$ decays can be used to determine
the modulus of $V_{ub}$, one of the smallest and least well known CKM
matrix elements. Compared to the determination with inclusive decays,
the extra kinematical constraints allow access to a larger part of the
lepton-momentum spectrum, resulting in smaller extrapolation
uncertainties. Experimentally, the main difficulty for the
observation of $b \rightarrow u \ell \nu$ signal events is the large
background from $b \rightarrow c \ell \nu$ events. Because
$|V_{ub}/V_{cb}|\approx 0.1$, the branching fractions of the exclusive
$b \rightarrow u \ell \nu$ decays ($\sim 10^{-4}$) are small compared
to those of the charmed semileptonic decays, which are of the order of
some percent. 

To extract the branching fraction $B \rightarrow \rho e \nu$ and
$|V_{ub}|$ requires use of hadronic form-factors which have to be
obtained from theory. In this analysis we use five different form
factor calculations: the two quark models ISGW2 \cite{isgw2} and
Beyer/Melikhov \cite{beyer98}, the lattice calculation by the UKQCD
group~\cite{ukqcd}, a model based on light cone sum rules
(LCSR~\cite{lcsr}), and a model based on heavy quark and $SU(3)$
symmetries (Ligeti/Wise~\cite{ligeti}).  The UKQCD and LCSR
calculations directly use QCD, whereas the quark models are more
phenomenological. All calculations use a particular value of
$q^2=(p_e+p_{\nu})^2$ as a normalization point. Usually, $q^2_{\rm
max}$ is used as this is the point where the hadronic system is least
disturbed. The LCSR result however is normalized at a lower value of
$q^2$.

\section{The \babar\ Detector and Data Set}

The data used in this analysis were collected with the \babar\
detector~\cite{nim} at the \pep2\ $\epem$ storage ring~\cite{pepii}.
The integrated luminosity of the sample is $50.5\invfb$ taken
at the \FourS\ mass (``on-resonance''), corresponding to $55.2$
million $B\bar{B}$ meson pairs. An additional $8.7\invfb$ of data were
taken $40\mev$ below the \FourS\ resonance (``off-resonance'').

\pep2\ is an \epem\ collider operated with asymmetric beam energies,
producing a boosted ($\beta\gamma = 0.55$) \FourS\ along the collision
axis.  \babar\ is a solenoidal detector optimized for the asymmetric
beam configuration at PEP-II. Charged particle (track) momenta are
measured in a tracking system consisting of a 5-layer, double-sided,
silicon vertex tracker (SVT) and a 40-layer drift chamber (DCH) filled
with a gas mixture of helium and isobutane, both operating within a
$1.5\,{\rm T}$ superconducting solenoidal magnet. Photon candidates
are selected as local maxima of deposited energy in an electromagnetic
calorimeter (EMC) consisting of 6580 CsI(Tl) crystals arranged in
barrel and forward endcap subdetectors. Particle identification is
performed by combining information from ionization measurements
(\dedx) in the SVT and DCH, and the Cherenkov angle $\theta_c$
measured by a detector of internally reflected Cherenkov light
(DIRC). The DIRC system is a unique type of Cherenkov detector that
relies on total internal reflection within the radiating volumes
(quartz bars) to deliver the Cherenkov light outside the tracking and
magnetic volumes, where the Cherenkov ring is imaged by an array of
$\sim 11000$ photomultiplier tubes. The detector is surrounded
by an instrumented flux-return (IFR).

\section{Event Selection}
\label{selection}

In this section, we describe the selection of the exclusive
semileptonic decays \brhoenu,\; \brhochgenu,\; \bomegaenu,\;
\bpienu,\; and \bpichgenu\; (with $\rho^0 \rightarrow \pi^+\pi^-$,
$\rho^\pm \rightarrow \pi^0\pi^\pm$ and $\omega \rightarrow
\pi^0\pi^+\pi^-$). The charge conjugate decays are implied
throughout. Our analysis strategy is similar to one used by
CLEO~\cite{dlange}. The analysis is optimized for $B \rightarrow \rho
e \nu$ decays; the $\pi$ and $\omega$ modes are included because of
the crossfeeds into the $\rho$ modes. Isospin and quark model
relations are used to effectively measure only two branching
fractions, one for $B \rightarrow \rho e \nu$ and one for $B
\rightarrow \pi e \nu$. This is described in section
\ref{fitMethod}. This analysis uses only electrons and not muons
because the background contribution from fake leptons is much lower in
the case of electrons. We reconstruct three kinematic variables which
are used in the fit to extract the signal yields. These are the
electron energy $E_{\rm lept}^{\rm CM}$, the invariant hadronic mass
$M_{\pi \pi (\pi)}$ (for the $\rho$ and $\omega$ modes), and the
difference between the reconstructed and expected $B$ meson energy
($\Delta E \equiv E_{\rm hadron} + E_{\ell} + \vert {\vec{p}}_{{\rm
miss}} \vert c - E_{\rm beam}$) in the center-of-mass (CM) system.

Two electron energy regions are considered: $2.0 \le
E^{\rm CM}_{\rm lept} < 2.3 \mbox{ GeV}$ (LOLEP), and $2.3 \le
E^{\rm CM}_{\rm lept} < 2.7 \mbox{ GeV}$ (HILEP).  The HILEP region is most
sensitive to the signal because the $b \rightarrow c e \nu$ events are
almost completely suppressed; the largest background source here is
from continuum $e^+e^- \rightarrow q\bar{q}$ 
events. Real data taken below the $\Upsilon(4S)$ mass, which includes
$e^+e^- \rightarrow e^+e^-(\gamma)$ events, is used for
the continuum subtraction.
In the LOLEP region, $b\rightarrow c$ decays
dominate and provide the normalization of the background at higher
electron energies.

Hadronic events are selected based on track multiplicity and event
topology. The tracks must have at least 12 hits in the drift
chamber. We also require that the impact parameter of the track along
and transverse to the beam direction must be less than 3 cm and 1 cm,
respectively. In addition, the transverse momentum must be greater
than $0.1\mbox{ GeV}/c$. Clusters in the electromagnetic calorimeter
of \babar\ that are not associated to any tracks must have an energy
greater than $30\mbox{ MeV}$ to be considered as photons. In addition,
the lateral moment of the shower energy distribution~\cite{LAT}
must be smaller than $0.8$. We select events with either $N_{\rm
tracks} \ge 5$ or ($N_{\rm tracks} \ge 4$ and $N_{\rm photons} \ge
5$). The $B$ mesons are produced nearly at rest so their decay
products are distributed roughly uniformly in solid angle. In
contrast, continuum events have a much more collimated (jet-like)
event topology. They are suppressed by requiring the ratio of
Fox-Wolfram moments $H_2/H_0$ \cite{foxw} to be less than~0.4. This
requirement suppresses 55\% of the $e^+e^- \rightarrow q\bar{q}$
background and non-hadronic events, with a signal efficiency of
85\%. In addition a neural net is used for further suppression of
continuum events, as described below.
 
To identify electrons, we use a likelihood estimator, which uses
information from several \babar\ sub-detectors. The primary information is
the ratio of the calorimeter energy to the track momentum. We require that
the direction of the electron momentum is within the good calorimeter
acceptance $-0.72<\cos \theta_{e,{\rm lab}}<0.92$. The efficiency of this
selector is around $90\%$, with a pion misidentification rate of less
than $0.1\%$. We also reject electrons from $J/\psi$ decays (requiring
two electrons identified with the likelihood electron selector and $3.00
<M_{e^+e^-} < 3.14 \mbox{ GeV}/c^2$) and from photon conversions ($M_{e^+e^-} <
30\mbox{ MeV}/c^2$). 

To reconstruct the neutral $\rho$ meson we combine two oppositely
charged tracks, and for the case of the charged $\rho$ 
a track and a $\pi^0$. The $\pi^0$ mesons are reconstructed from two photons with an
invariant mass $120 < M_{\gamma\gamma} < 145\mbox{ MeV}/c^2$
corresponding to $\pm 2\ \sigma$ from the nominal $\pi^0$ mass on average. To
suppress combinatorial backgrounds we require that the pion with the
highest momentum must have $p_{\pi}^{\rm CM}>400\mbox{ MeV}/c$ and the
other pion must satisfy $p_\pi^{\rm CM}>200\mbox{ MeV}/c$. For the
$\omega$, we combine two oppositely charged tracks with a $\pi^0$. The
$\omega$ invariant mass is measured within $\pm 80\mbox{
MeV}/c^2$ of the nominal $\omega$ mass~\cite{pdg2000}.
This includes a side band region below and
above the nominal $\omega$ mass. To suppress combinatorial backgrounds
we require $p_\pi^{\rm CM}>100\mbox{ MeV}/c$ for each of the three pions.
The charged tracks used to reconstruct the $rho$, $omega$, or $pi^\pm$ mesons 
must not have been identified as kaons.

In the following discussion all variables are taken in the 
center-of-mass frame. The neutrino is reconstructed from the missing momentum:
\begin{equation}
\vec{p}_{\rm miss} = - \sum_{\rm tracks}{\vec{p}_i} - \sum_{\rm photons}{\vec{p}_i}\;,
\end{equation}
where the sums run over all reconstructed and accepted tracks and
photons in the event. We then take
$(E_{\nu},\vec{p}_\nu)=(|\vec{p}_{\rm miss}|c,\vec{p}_{\rm miss})$.

\begin{figure}[tb]
\begin{center}
\input{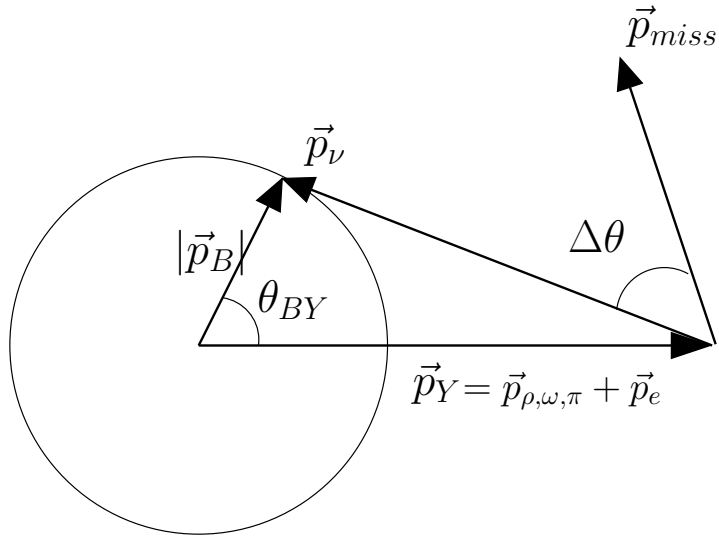}
\unitlength=1cm
\put(-3.4,1.85){\Large $= \vec{p}_{\rho,\omega,\pi}+ \vec{p}_e$}
\caption{Angles $\theta_{BY}$ and $\Delta \theta$ defined in the $\Upsilon(4S)$ frame.}
\label{fig:thetaby}
\end{center}
\end{figure}

A $B$-meson decay consistent with the signal modes is reconstructed
using the constraints $E_{B} = E_{\rm beam}$ and $(p_B - p_Y)^2 = 0$,
where $Y$ is the $(\rho,~\omega,~\pi) + e$ system. A useful quantity
for testing consistency is the angle between the $B$ momentum
direction and that of the reconstructed $Y$ system, see Fig.~\ref{fig:thetaby},
\begin{equation}
\cos\theta_{BY}={2 E_B E_{Y} - (M_B^2  + M_{Y}^2) c^4 
                      \over 2 |\vec{p}_B| |\vec{p}_{Y}| c^2}\;.
\label{eq:cosby}
\end{equation}
Background tends to have non-physical values of
$\cos\theta_{BY}$. For the signal, small extensions of $|\cos\theta_{BY}|>1$
are allowed because of detector resolution. We therefore require
\begin{equation}
|\cos \theta_{BY}| < 1.1\; .
\end{equation}
The efficiency of this requirement is almost 100\% for the signal and it
rejects more than 60\% of the $b \rightarrow c e \nu$ and 80\% of
the continuum background.

We also compare the direction of the missing momentum $\vec{p}_{{\rm
miss}}$ with that of the neutrino momentum inferred from
$\vec{p}_{\nu} = \vec{p}_{B}-\vec{p}_{Y}$. The latter is known to
within an azimuthal ambiguity about the $\vec{p}_{B}$ direction
because the magnitude, but not the direction, of the $B$ meson
momentum is known. We use the smallest possible angle
$\Delta\theta_{\rm min}$ between the two directions, which is obtained
when the momenta of $\vec{p}_Y$, $\vec{p}_{\nu}$, and $\vec{p}_{\rm
miss}$ are in the same plane, see Fig.~\ref{fig:thetaby}. We use the
requirement
\begin{equation}
0.8< \cos \Delta\theta_{\rm min}\le 1.0\;.
\end{equation} 
This has been optimized using a Monte Carlo simulation, in such a way as
to minimize the relative error on the measured branching fraction.

In addition, we require $\vert \cos \theta_{{\rm miss}}
\vert<0.9$ where $\theta_{\rm miss}$ is the angle between $\vec{p}_{\rm
miss}$ and the beam axis. This cut rejects events with missing high
momentum particles close to the beam axis.

The continuum $e^+e^- \rightarrow q\bar{q}$, where $q=u,d,s,c$, is an
important background at high electron energies, where we are most
sensitive to the signal. To reject these events, we use a neural net
with 14 event shape variables such as the track and cluster energies
in nine cones around the electron-momentum axis. The optimized cut on
the neural net output suppresses more than 90\% of the continuum,
after all other requirements have been applied, in the HILEP region. The
selection efficiency on the signal is 60\%.

After all the above criteria, there can still be several candidates
per event. This follows from the large width of the $\rho$ and also
because we are reconstructing five different decay modes. To avoid
statistical difficulties related to large numbers of combinations, we
choose one candidate per event, namely the one with a reconstructed
total momentum closest to the B meson momentum:
\begin{equation}
|\vec{p}_{\rm hadron} + \vec{p}_e + \vec{p}_{\rm miss}|\;\; {\rm closest \; to}\;\; |\vec{p}_B|.
\end{equation}
The efficiency of this selection for the signal is close to 85\% .

At each step of the selection procedure the data distributions agree
well with their Monte Carlo simulation.

The total efficiencies in the HILEP region are $4.21\%$ for
the $B^+ \rightarrow \rho^0 e^+ \nu$ mode and $3.31\%$ for
the $B^0 \rightarrow \rho^- e^+ \nu$ mode. These efficiencies are
determined using the ISGW2 form-factors. They are defined here as the
number of HILEP signal events that pass all selection criteria and are
reconstructed in the specified channel divided by the number of all
generated events (of all electron momenta) for this channel. 
The signal and crossfeed efficiencies for all channels are listed
in Table~\ref{tab:effe2}.

\begin{table}[htb]
\caption[HILEP selection efficiencies for the signal modes]{Selection 
efficiencies for the modes $B^+ \rightarrow \rho^0 e^+ \nu$, $B^0
\rightarrow \rho^- e^+ \nu$, $B^+ \rightarrow \omega e^+ \nu$, $B^+
\rightarrow \pi^0 e^+ \nu$, and $B^0 \rightarrow \pi^- e^+ \nu$ in the HILEP 
region. These efficiencies have been determined using simulated data (ISGW2
model). They are defined as the number of selected events passing all cuts
divided by the number of all generated events (in the full lepton energy 
range) for the specified channel.}
\begin{center}
\begin{tabular}{l|c|c|c|c|c} \hline \hline
  & \multicolumn{5}{c}{{\bf Reconstruction Efficiency (\%)}} \\ \cline{2-6}
  {\bf Generated}          &  $B^+ \rightarrow \rho^0 e^+ \nu$ & $B^0 \rightarrow \rho^- e^+ \nu$ & $B^+ \rightarrow \omega e^+ \nu$ &  $B^+ \rightarrow \pi^0 e^+ \nu$ & $B^0 \rightarrow \pi^- e^+ \nu$ \\ \hline
$B^+ \rightarrow \rho^0 e^+ \nu$ & $4.21$ & $0.99$ & $0.30$ & $0.58$ & $1.35$\\
$B^0 \rightarrow \rho^- e^+ \nu$ & $1.04$ & $3.31$ & $0.31$ & $1.08$ & $1.32$\\
$B^+ \rightarrow \omega e^+ \nu$ & $1.97$ & $1.55$ & $1.57$ & $0.94$ & $0.88$\\
$B^+ \rightarrow \pi^0 e^+ \nu$  & $0.23$ & $0.41$ & $0.07$ & $1.28$ & $0.40$\\
$B^0 \rightarrow \pi^- e^+ \nu$  & $0.42$ & $0.47$ & $0.05$ & $0.33$ & $1.63$\\
\hline \hline
\end{tabular}
\end{center}
\label{tab:effe2}
\end{table}

\section{Fit Method}
\label{fitMethod}

We have performed a binned maximum-likelihood fit of the
two-dimensional distribution ($M_{\pi \pi (\pi)}$, $\Delta E$)
simultaneously in the two electron energy ranges (LOLEP, HILEP) and
the decay modes \brhoenu,\; \brhochgenu,\; and \bomegaenu. For the $B
\to \rho e \nu$ modes, the data are divided into $10 \times 10$ bins
over the ($M_{\pi\pi}$, $\Delta E$) region $0.25 \leq M_{\pi\pi} \leq
2.00\mbox{ GeV}/c^2$ and $|\Delta E| \leq 2\mbox{ GeV}$. The bin size for
the fit is thus $175 \mbox{ MeV}/c^2$ in $M_{\pi\pi}$ and $400\mbox{
MeV}$ in $\Delta E$. For the $\omega$ channel, we use 5 bins in the
range $782 \pm 80 \mbox{ MeV}/c^2$ and 10 bins in $|\Delta E|\leq
2\mbox{ GeV}$. The modes \bpienu\; and \bpichgenu\; are also included
to model the crossfeeds into the other signal channels; for these
modes only $\Delta E$ is used as a fit variable.

Our fit includes contributions from the signal modes, other $b \to u e
\nu$ decays, $b \to c e \nu$ decays, continuum, and a contribution from
misidentified electrons. For the signal and backgrounds coming from
other $b \to u e \nu$ and $b \to c e \nu$ decays, Monte Carlo
simulation provides the shapes of the distributions. The decays $B \to
D^* \ell \nu$ have been simulated using heavy quark effective theory
(HQET \cite{HQET}). The modes $B \to D^{*} \pi \ell \nu$ are simulated according
to the Goity-Roberts model~\cite{goity}. Resonant $b \to u \ell \nu$
downfeed modes are implemented according to the ISGW2
model. Non-resonant $b \to u \ell \nu$ modes are implemented according
to a model by Fazio and Neubert \cite{Xu}.

Isospin and quark model relations are used to constrain the relative
normalizations of $B^0 \rightarrow \rho^- e^+ \nu$, $B^+
\rightarrow \rho^0 e^+ \nu$, and $B^+ \rightarrow \omega e^+
\nu$ and therefore to reduce the number of free fit parameters:
\begin{eqnarray}
\Gamma(B^0 \rightarrow \rho^- e^+ \nu) & = & 2
\Gamma(B^+ \rightarrow \rho^0 e^+ \nu), \label{iso1}\\
\Gamma(B^+ \rightarrow \rho^0 e^+ \nu) & = &  
\Gamma(B^+ \rightarrow \omega e^+ \nu), \label{iso2} \\
\Gamma(B^0 \rightarrow \pi^- e^+ \nu) &=& 2
\Gamma(B^+ \rightarrow \pi^0 e^+ \nu).
\label{eq:isorho}
\end{eqnarray}
Isospin breaking effects are discussed in \cite{iso} and
\cite{dlange2}.  We assume that the isospin relations in
Eqs.~\ref{iso1} and \ref{iso2} are broken by not more than $3\%$. This
would have a negligible effect on our result and therefore we do not
include a corresponding systematic error. The isospin relations were
tested experimentally. We find that the isospin relations 
Eqs.~\ref{iso1} and \ref{iso2} are
consistent with the data within $1.3\;\sigma$ and $1.7\;\sigma$. 

We use the following 9 free parameters for the fit:
\begin{itemize}
\item[$\bullet$]  ${\cal B}(B^0 \rightarrow \rho^- e^+ \nu)$ (1 parameter);
\item[$\bullet$]  ${\cal B}(B^0 \rightarrow \pi^- e^+ \nu)$ (1 parameter);
\item[$\bullet$]  the scale factors of the $b \rightarrow u e \nu$ background in each electron energy
bin (2 parameters), that give the overall normalization of all 
$b \to u e  \nu$ modes that are not signal modes, relative to that expected
from the Monte Carlo simulation;
\item[$\bullet$]  the scale factors, one for each mode, that give the overall normalization of the
$b \to c e  \nu$ background relative to that expected
from the Monte Carlo simulation (5 parameters).
\end{itemize}

The maximum-likelihood fit method used in this analysis has been
described in \cite{barlow}. The fit takes into account the statistical
fluctuations not only of the on-resonance and off-resonance data but
also those of the Monte Carlo contributions. We have performed a toy
Monte Carlo check to verify the stability of the fit method and to
check the statistical error returned by the fit.

Whereas the signal modes \brhoenu, \brhochgenu, and \bomegaenu\; are
simulated with five different form-factors, all other $b \rightarrow u
e \nu$ modes (downfeed background) are simulated using the ISGW2
form-factor only.

\section{Fit Results}

\begin{table}
\caption[Summary of data yields for the $\rho$ modes]{Summary of data
yields for the \brhochgenu\; and \brhoenu\; modes with electron energies
between $2.3$ and $2.7\mbox{ GeV}$ (HILEP), and between $2.0$ and
$2.3\mbox{ GeV}$ (LOLEP). The yields presented in this table were
obtained using the ISGW2 form-factor. The downfeed background includes
all $B \rightarrow X_u e \nu$ modes except for $\rho$, $\omega$, and
$\pi$. The crossfeed signal contribution corresponds to events from
the other signal modes with $\rho^0$, $\omega$, or $\pi$ and is
constrained to the signal in the fit. All errors are statistical
only.}
\begin{center}
\begin{tabular}{l|cc|cc}\hline \hline
   & \multicolumn{2}{c|} \brhochgenu & \multicolumn{2}{c} \brhoenu \\
   & HILEP &  LOLEP & HILEP & LOLEP \\ \hline 
On-resonance yield              & 2302         & 39349           & 2213         & 40155 \\
Direct signal                   & 510 $\pm$ 63 & 718 $\pm$ 89    & 324 $\pm$ 40 & 440 $\pm$ 55 \\
Crossfeed signal                & 262 $\pm$ 32 & 538 $\pm$ 73    & 363 $\pm$ 42 & 725 $\pm$ 86 \\
Downfeed                        & 203 $\pm$ 55 & 2278 $\pm$ 403  & 226 $\pm$ 92 & 2435 $\pm$ 430 \\
$b \rightarrow c e \nu$         & 414 $\pm$  5 & 33859 $\pm$ 438 & 367 $\pm$ 5  & 34366 $\pm$ 458 \\
$e^+ e^- \rightarrow q \bar q$  & 917 $\pm$ 73 & 1928  $\pm$ 106 & 912 $\pm$ 73 & 2063 $\pm$ 110 \\
Fake electrons                  &  12 $\pm$  3 & 80    $\pm$   9 & 18  $\pm$  4 & 76 $\pm$ 9 \\ 
\hline \hline
\end{tabular}
\end{center}
\label{table:yieldsrhochg}
\end{table}

\begin{figure}[htb]
\begin{center}
\epsfig{file=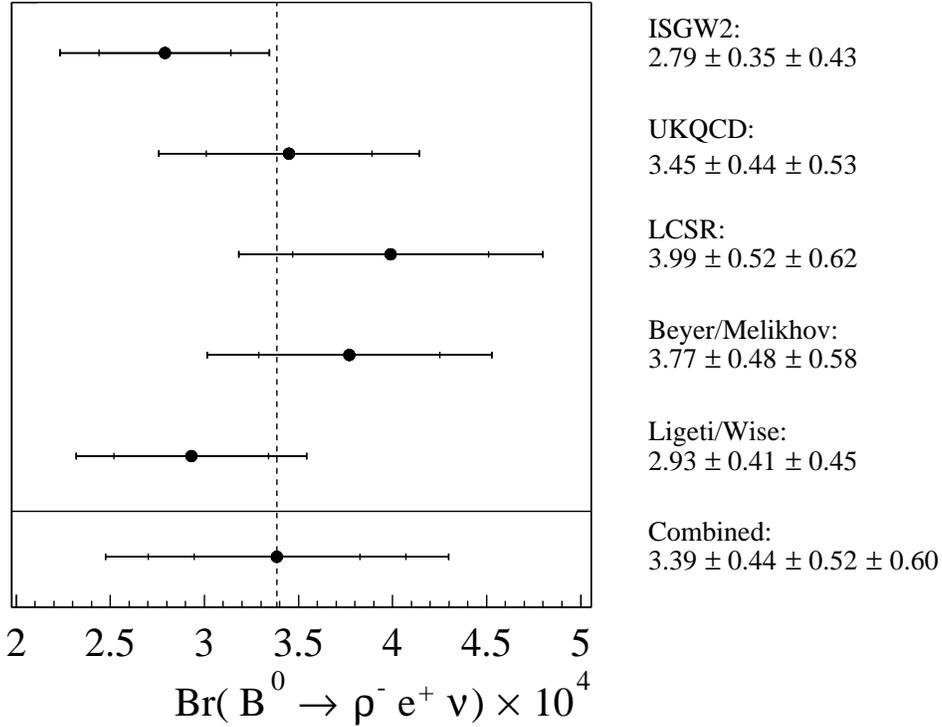,width=14cm}
\end{center}
\caption[\brhochgenu\; branching fraction for different form-factors]{The
\brhochgenu\; branching fraction results using the ISGW2, UKQCD, LCSR,
Beyer/Melikhov, and Ligeti/Wise form-factors. The errors shown are
statistical, systematic, and (only in case of the combined result)
theoretical, successively combined in quadrature. The combined central
value is determined by taking the unweighted mean of all form-factor
results. The statistical and systematic errors of the combined result
are determined by taking the means of the relative errors of each
individual result, and its theoretical error is taken to be one half of
the full spread of the results.}
\label{fig:brrho}  
\end{figure}

The signal yields extracted from the binned maximum-likelihood fit in
the HILEP region are 324~$\pm$~40 \brhoenu\; events and 510~$\pm$~63
\brhochgenu\; events, based on the ISGW2 calculation. The composition
of events for the $B^0 \rightarrow \rho^- e \nu$ and $B^+
\rightarrow \rho^0 e^+ \nu$ channel is shown in
Table~\ref{table:yieldsrhochg}. The isospin-constrained results for
the five different form-factors are shown in Fig.~\ref{fig:brrho}. A
$\chi^2$ test has been performed to check the quality of the fit.
Bins in sparsely populated regions have been combined before the
$\chi^2$ calculation. For ISGW2, we obtain $\chi^2=91$ for $N_{\rm
dof} = 93$, which corresponds to a $p$-value of $0.52$, and similarly good fit
quality for the other four form-factor calculations.

The five fit parameters describing the $b \rightarrow c$ backgrounds
agree with the Monte Carlo expectations within $9\%$ on average. The
two parameters describing the $b \rightarrow u$ downfeed background in
LOLEP and HILEP agree to better than $1.5\ \sigma$ and $1.2\ \sigma$.
The fit result for the $\pi$ modes is 
${\cal B}(B^0 \rightarrow \pi^- e^+ \nu)= (1.87 \pm 0.56) \times 10^{-4}$ 
for the ISGW2 calculation.

\begin{figure}
  \begin{center}
    \subfigure[$M_{\pi\pi}$ (LOLEP)]{\epsfig{file=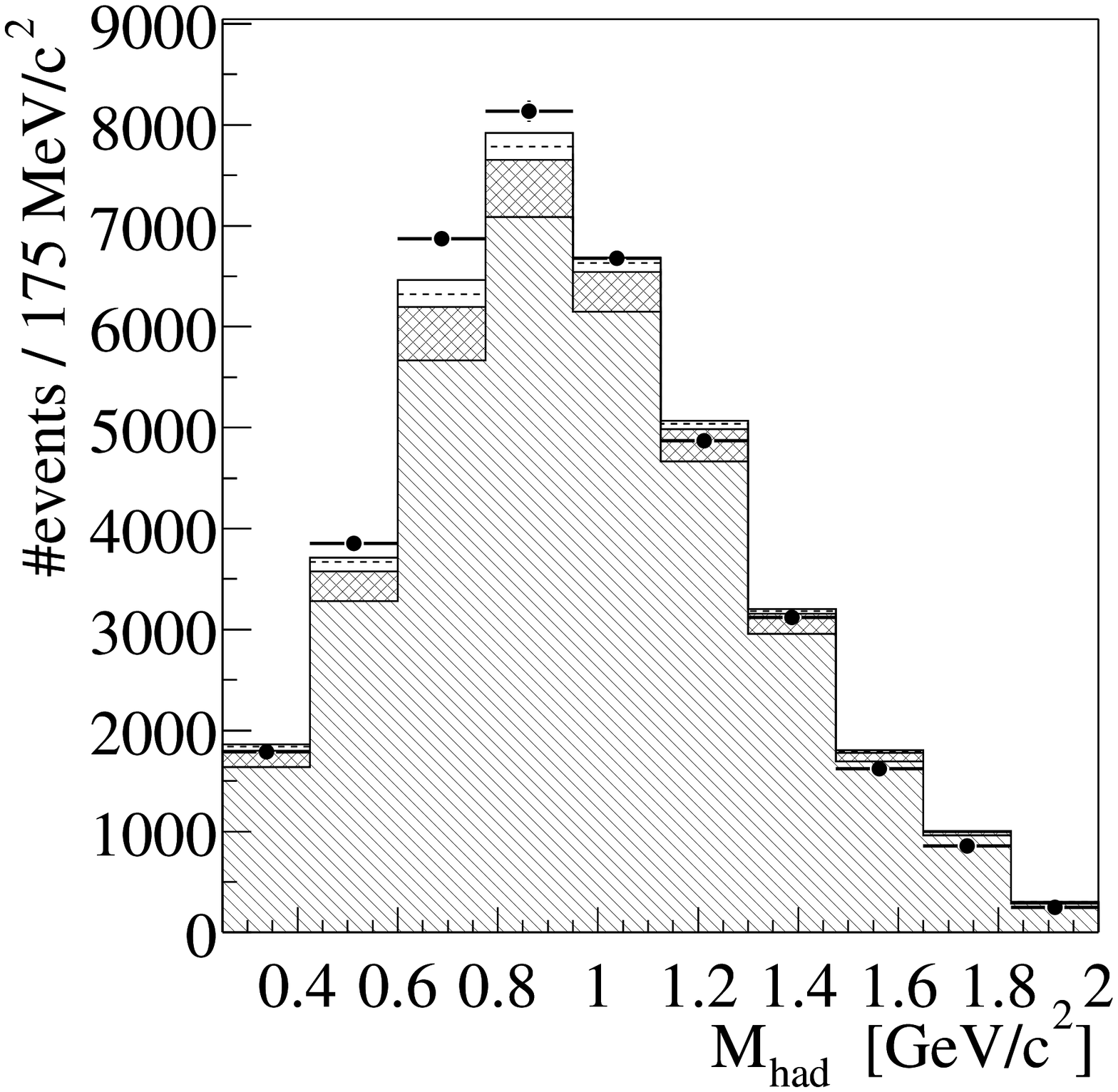,width=7cm}}
    \subfigure[$M_{\pi\pi}$ (HILEP)]{\epsfig{file=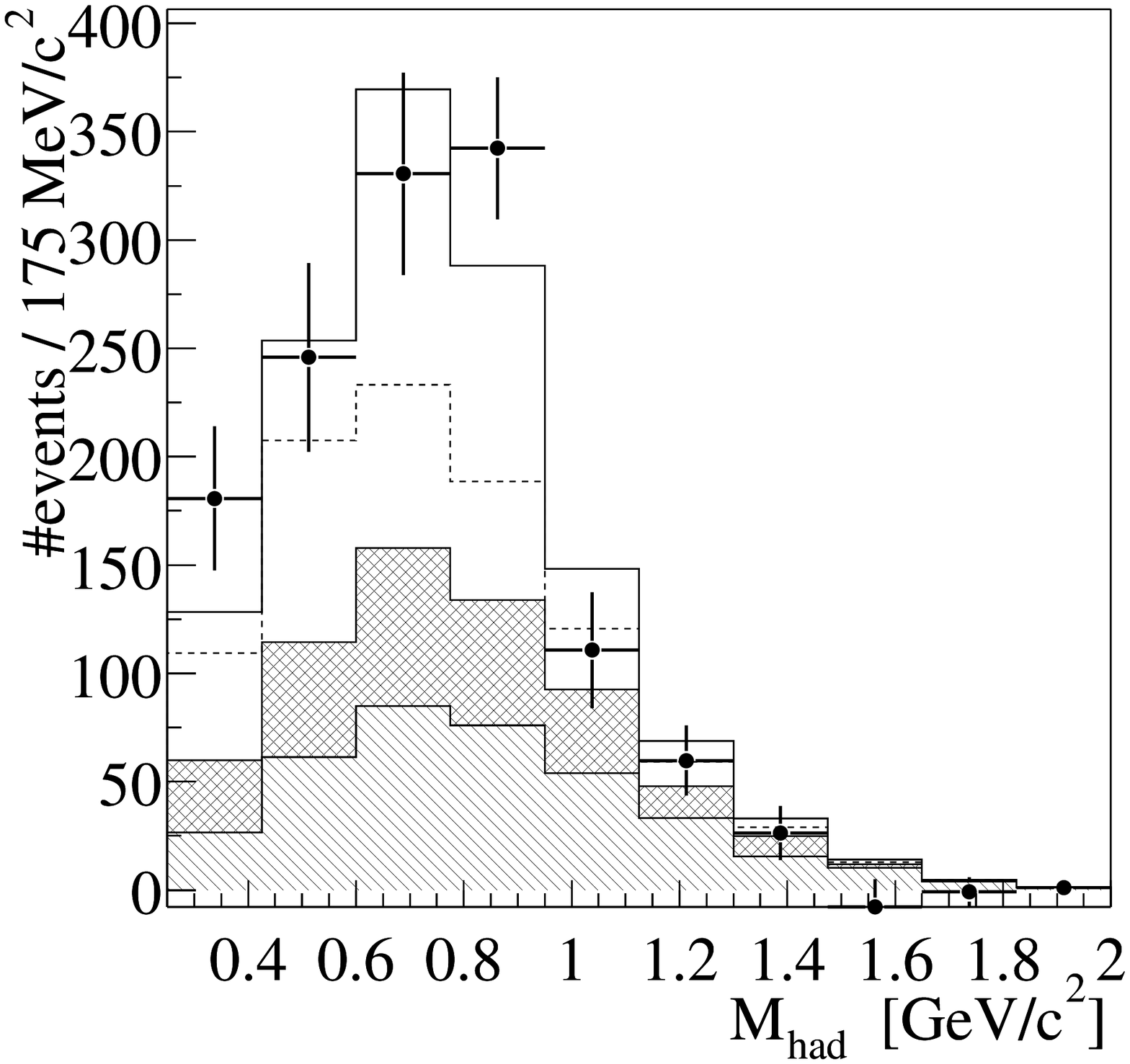,width=7cm}}
    \subfigure[$\Delta E$ (LOLEP)]{\epsfig{file=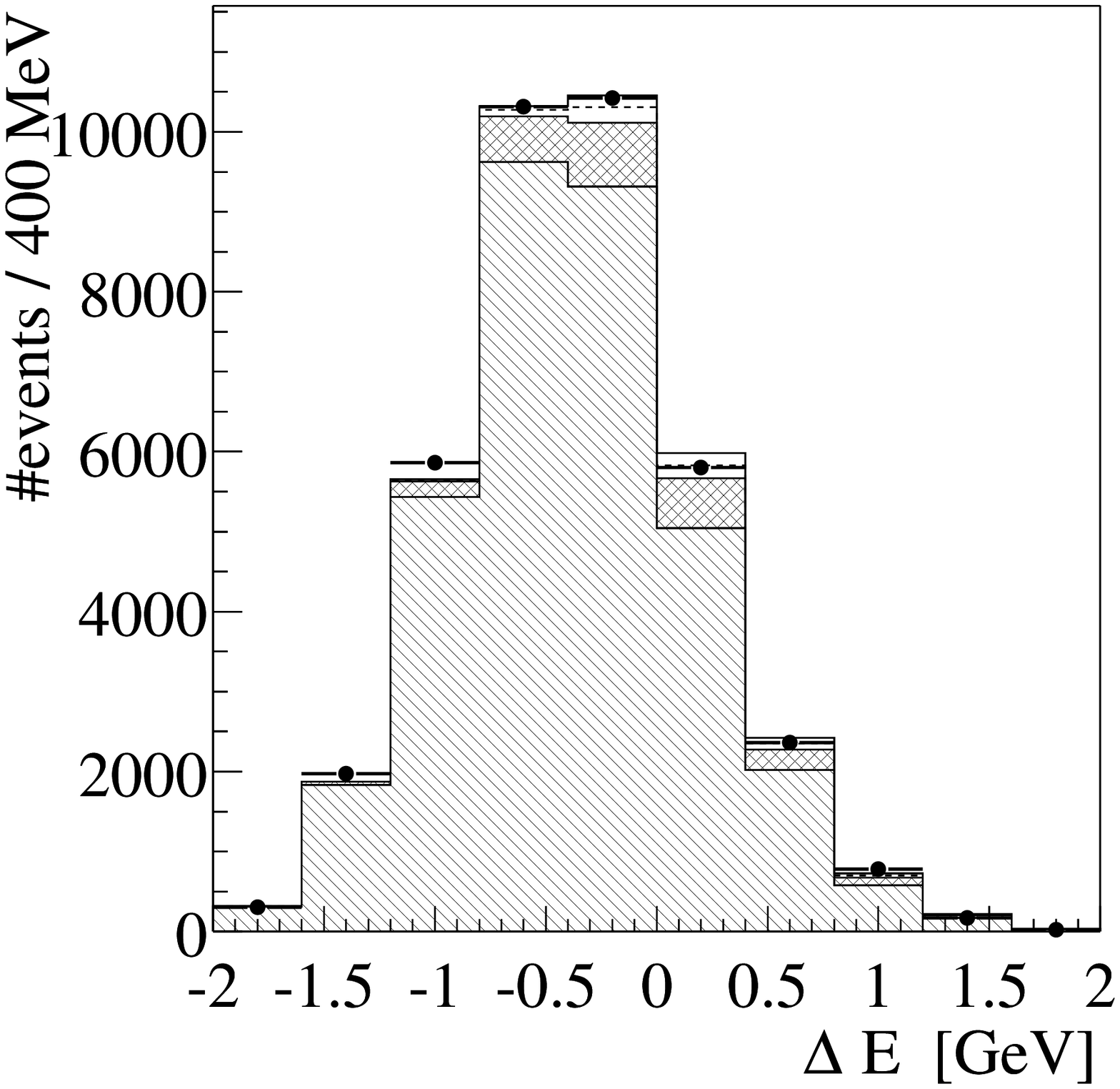,width=7cm}}
    \subfigure[$\Delta E$ (HILEP)]{\epsfig{file=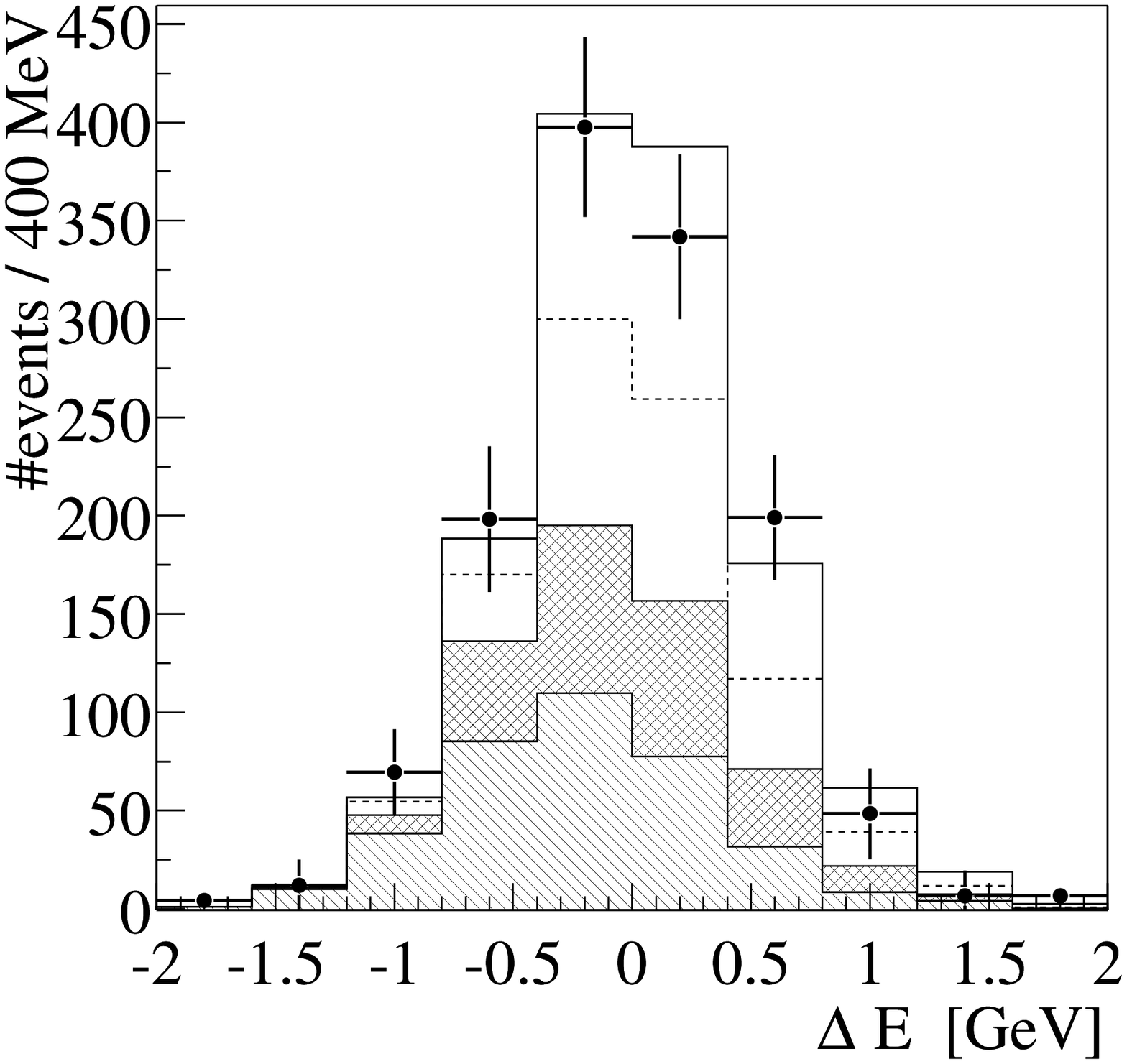,width=7cm}}
\caption{Continuum-subtracted projections of the ISGW2 fit result for the \brhoenu\ channels in
the LOLEP and HILEP electron energy regions; the contributions are the
direct and crossfeed components of the signal (unhatched region, above
and below the dashed line, respectively); the background from $b\to u
e \nu$ other than $B \rightarrow \rho e \nu$ and $B \rightarrow \omega
e \nu$ modes (double-hatched region); the background from $b\to c e
\nu$ and other backgrounds (single-hatched region).}
\label{fig:resrho} \end{center}
\end{figure}

\begin{figure}
  \begin{center}
    \subfigure[$M_{\pi\pi}$ (LOLEP)]{\epsfig{file=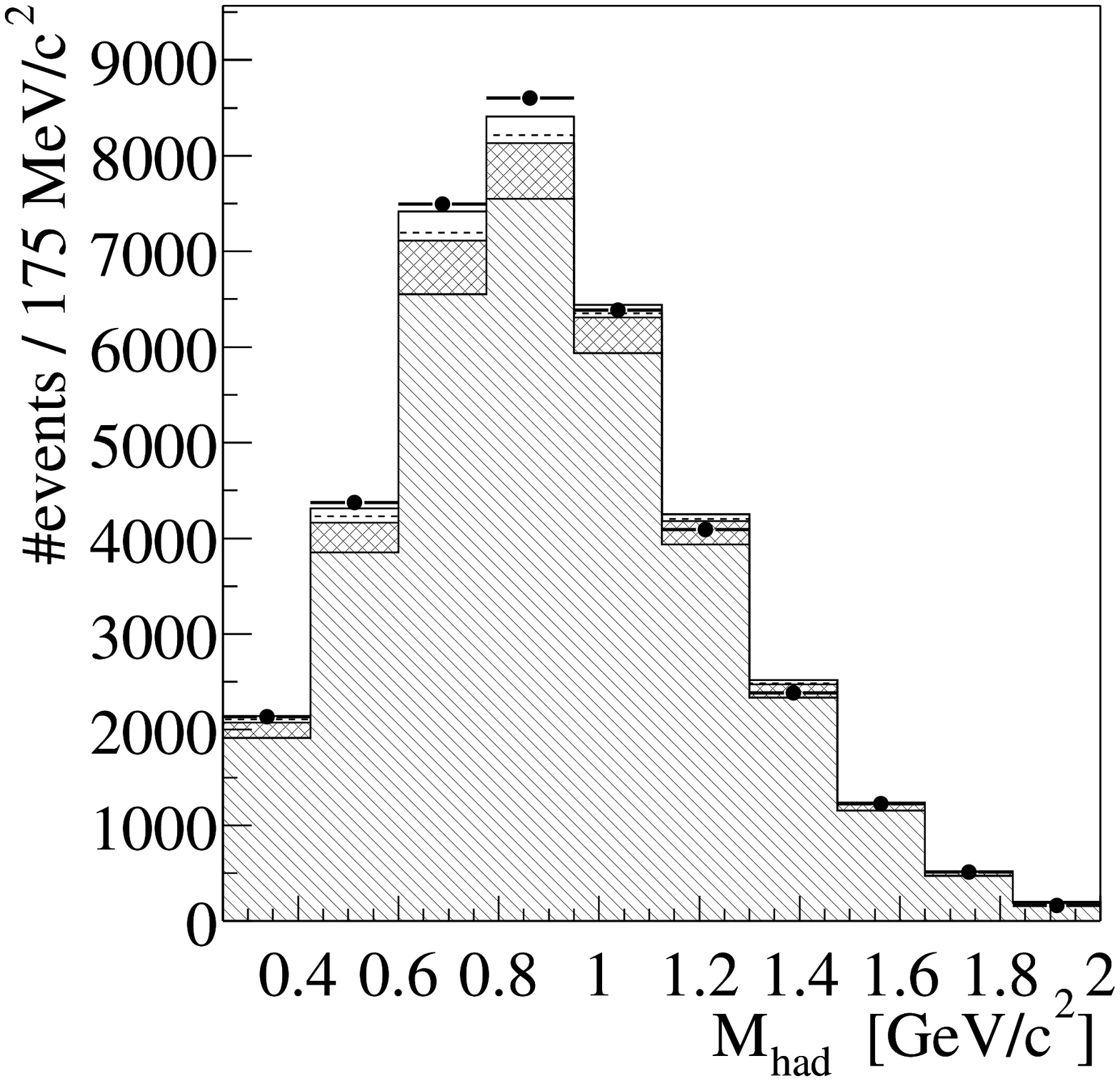,width=7cm}}
    \subfigure[$M_{\pi\pi}$ (HILEP)]{\epsfig{file=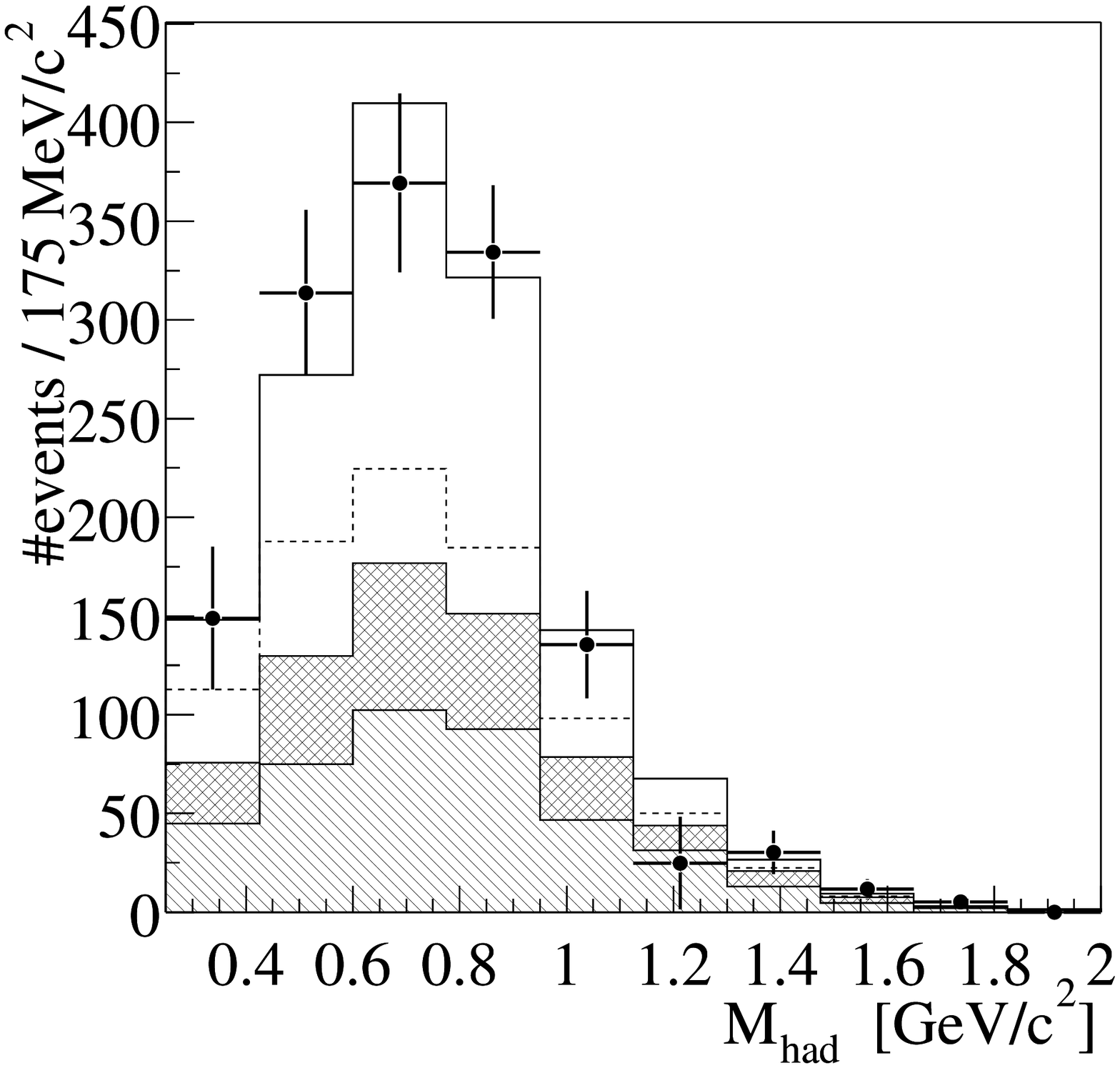,width=7cm}}
    \subfigure[$\Delta E$ (LOLEP)]{\epsfig{file=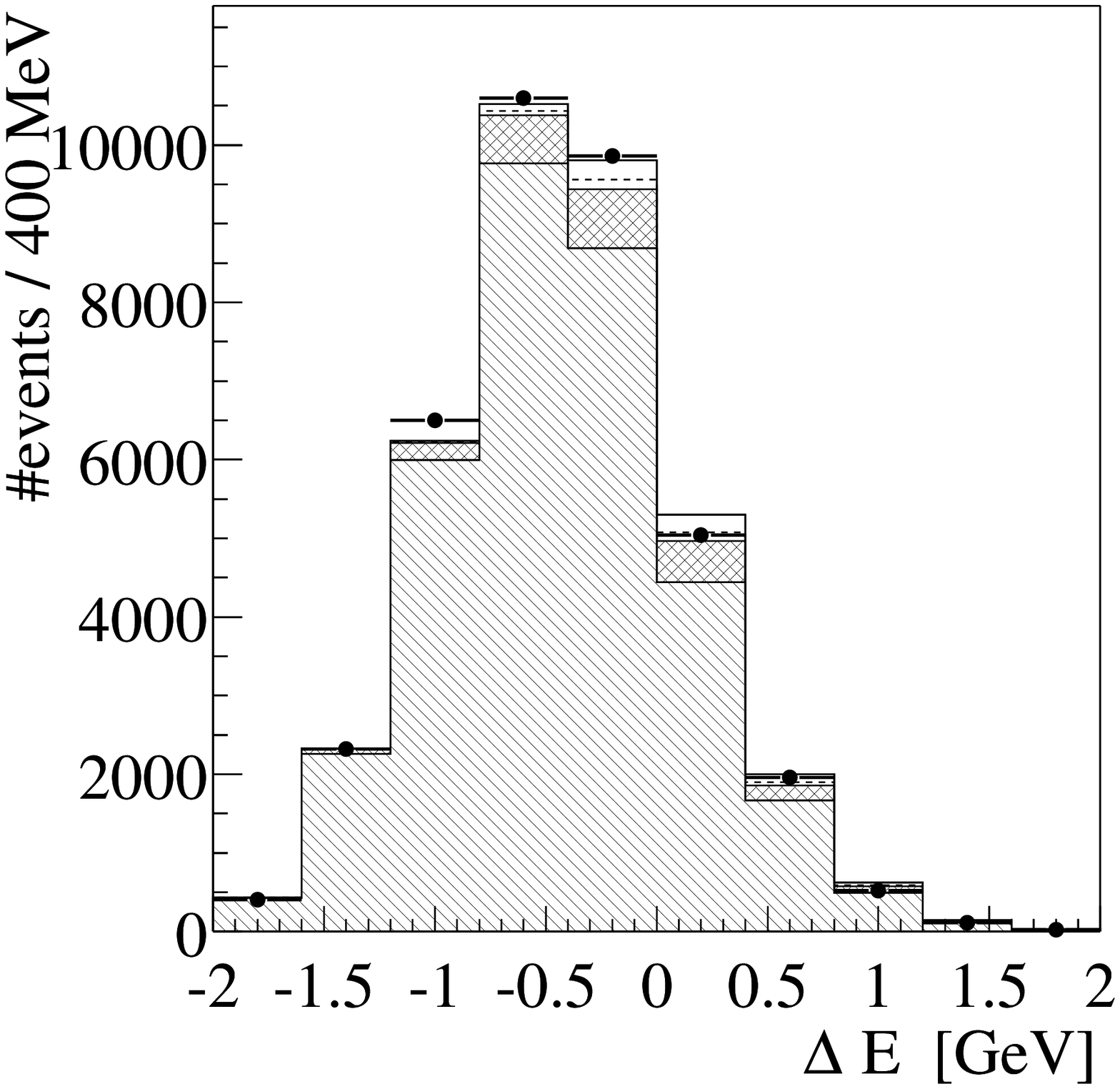,width=7cm}}
    \subfigure[$\Delta E$ (HILEP)]{\epsfig{file=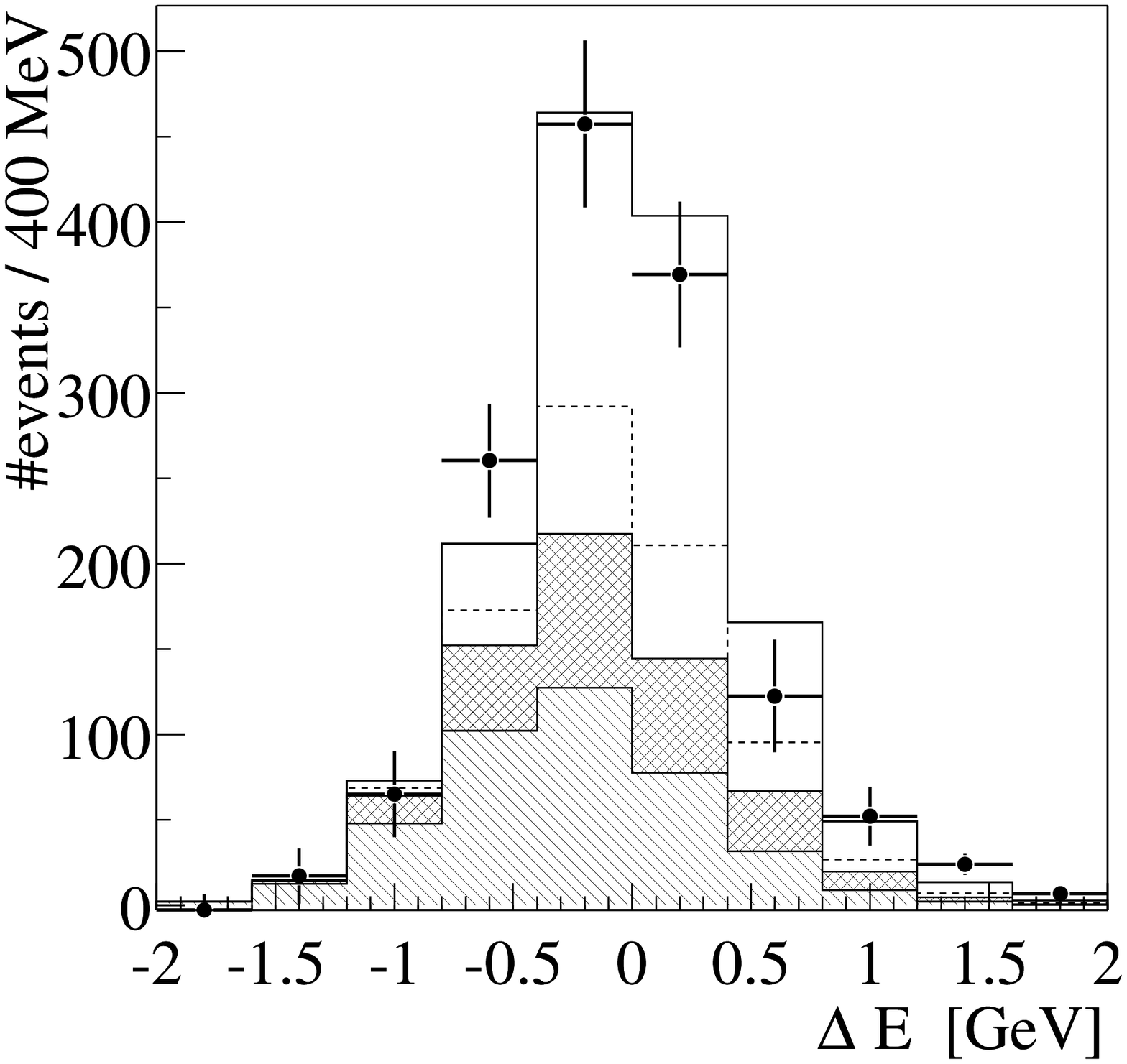,width=7cm}}
    \caption{Continuum-subtracted projections of the ISGW2 fit result for the \brhochgenu\ 
channels in the LOLEP and HILEP electron energy regions; the
contributions are the direct and crossfeed components of the signal
(unhatched region, above and below the dashed line, respectively); the
background from $b\to u e \nu$ other than $B \rightarrow \rho e \nu$
and $B \rightarrow \omega e \nu$ modes (double-hatched region); the
background from $b\to c e \nu$ and other backgrounds (single-hatched
region).}  \label{fig:resrhochg} \end{center}
\end{figure}

The projections of the ISGW2 fit result for the two electron energy
bins after subtraction of the continuum contribution are shown in
Figs.~\ref{fig:resrho} and \ref{fig:resrhochg}. Good agreement between
the data and fit result is seen in each of these figures. The fits for
the other four form-factor calculations show similar agreement.

\section{Systematic Errors}

The summary of all systematic errors on the branching fraction that
have been considered is shown in Table~\ref{systematicErrors}. The
total systematic error is taken as the quadratic sum of all individual
errors. The fluctuations due to finite Monte Carlo statistics are
included in the statistical error and not in the systematic error.

The largest single systematic error comes from the uncertainty in the
shape of the downfeed background. We use the ISGW2 model to describe
resonant downfeed modes, and a model by Fazio and Neubert \cite{Xu}
for non-resonant modes. The fraction of non-resonant events in the
downfeed background is varied from $0\%$ to $68\%$ to estimate this
systematic uncertainty. The composition of the resonant $b \rightarrow
u$ downfeed component has been varied by changing the branching
fraction for individual resonances by $\pm 50\%$, and keeping the
total rate constant.

We have also varied the most important selection requirements of this 
analysis within a reasonable range and have changed our fit method
(fitting with only four channels, or without the LOLEP region, or with 
different binnings). Most variations seen are within $1\sigma$
of the expected statistical variation, some are close to $2\sigma$.
To be conservative we assign a systematic error corresponding
to half the largest variations seen. This corresponds to the
last two systematic errors quoted in Table \ref{systematicErrors}.

\begin{table}[!h]
\caption[Summary of all systematic errors]{Summary of all contributions to the systematic error on the branching fraction for $B \rightarrow \rho e \nu$.}
\begin{center}
\begin{tabular}{l|c} \hline \hline
Error contribution                         & $\delta{\cal B}_{\rho}/{\cal B}_{\rho}$ (\%) \\ 
\hline \hline
Tracking Efficiency                        & $\pm 5$\\
Tracking Resolution                        & $\pm 1$\\
Photon/$\pi^0$ Efficiency                  & $\pm 5$\\
Photon/$\pi^0$ Energy Scale                & $\pm 3$\\
\hline
$b\rightarrow c$ Background Composition    & ${}^{+1.4}_{-1.7}$ \\
Resonant $b\rightarrow u$ Background Composition & ${}^{+6}_{-4}$ \\
Non-Resonant $b\rightarrow u$ Background   & $\pm 9$\\
\hline
B Lifetime                                 & $\pm 1$ \\
B Counting                                 & $\pm 1.6$ \\
Fake Electrons                             & $<\pm 1$ \\
Electron Id                                & $\pm 2$ \\
$f_\pm/f_{00}$                             & $<\pm 1$ \\
Data Selection                             & $\pm 6$ \\
Fit Method                                 & ${}^{+4}_{-6}$ \\ \hline \hline
{\bf Total Systematic Error}               & $\pm 15.5$ \\
\hline \hline
\end{tabular}
\end{center}
\label{systematicErrors}
\end{table}

\section{Extraction of {\boldmath $|V_{ub}|$}}

\begin{table}[!tb]
\caption[$\tilde{\Gamma}_{{\rm thy}}$ of the form-factors]{ 
$\tilde{\Gamma}_{{\rm thy}}$ predicted by various form-factor calculations.}
\begin{center}
\begin{tabular}{lcc} \hline \hline
Form-factor      & $\tilde{\Gamma}_{{\rm thy}}$ (ps$^{-1}$) 
  & Estimated error on $\tilde{\Gamma}_{{\rm thy}} (\%)$ \\ \hline
ISGW2            & 14.2 & $\pm 50$ \\
LCSR             & 16.9 & $\pm 32$ \\
UKQCD            & 16.5 & ${}^{+21}_{-14}$ \\
Beyer/Melikhov   & 16.0 & $\pm 15$ \\ 
Ligeti/Wise      & 19.4 & $\pm 29$ \\  \hline \hline
\end{tabular}
\end{center}
\label{tab:gamthy}
\end{table}

\begin{figure}[htb]
\begin{center}
\epsfig{file=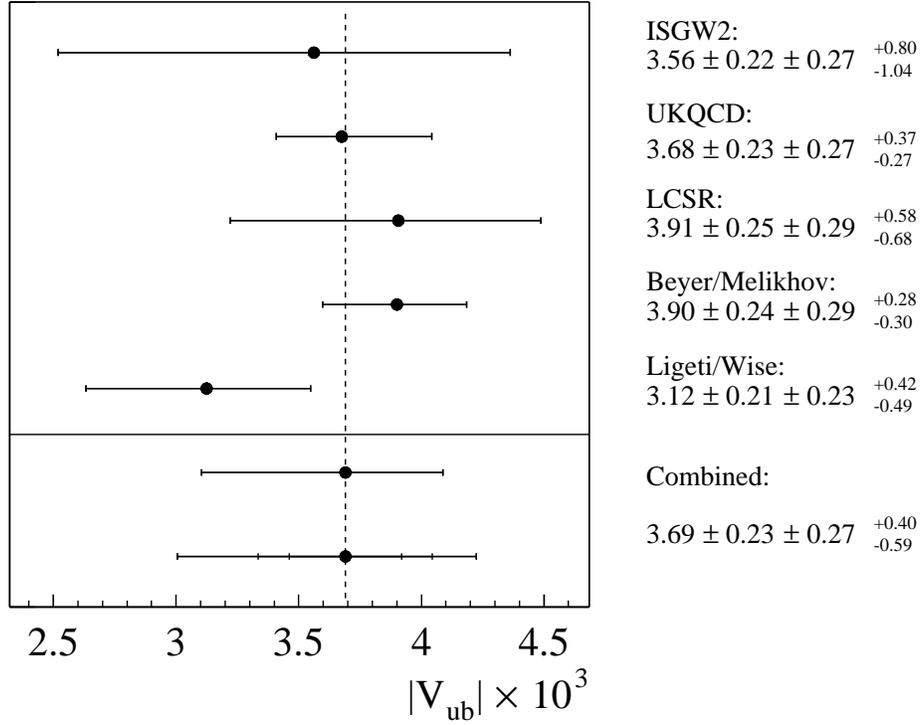,width=14cm}
\end{center}
\caption[$|V_{ub}|$ for different form-factors]{ $|V_{ub}|$ determined
using the ISGW2, UKQCD, LCSR, Beyer/Melikhov, and Ligeti/Wise
form-factors. The results for each form-factor are drawn with
theoretical error bars only. The combined central value is determined
by taking the mean of all form-factor results, weighted by their
individual theoretical error. The theoretical error of the combined
result is taken to be one half of the full spread of results,
including the errors. The combined result is also drawn with
statistical, systematic, and theoretical errors successively added in
quadrature. The statistical and systematic errors of the combined
result are determined by taking the mean of the relative errors of
each individual result. In addition we give, on the right side of the
figure, all five results with their statistical, systematic, and
theoretical errors.}
\label{fig:vub}   
\end{figure}

The CKM matrix element $|V_{ub}|$ can be obtained from the branching
fraction ${\cal B}(B^0 \to \rho^- e^+ \nu)$ using
\begin{equation}
\vert V_{ub} \vert = 
\sqrt{ {   {\cal B}(B^0\rightarrow \rho^- e^+ \nu) \over  
{\tilde{\Gamma}_{{\rm thy}} \tau_{B^0}}}}\;,
\label{eq:vub}
\end{equation}
where $\tilde{\Gamma}_{{\rm thy}}$ is the predicted form-factor
normalization. Values of $\tilde{\Gamma}_{{\rm thy}}$ and theoretical
errors for each form-factor calculation are given in
Table~\ref{tab:gamthy}. The calculations quote errors on
$\tilde{\Gamma}_{{\rm thy}}$ between $15\%$ and $50\%$. We use
$\tau_{B^0} = 1.548 \pm 0.032$~ps~\cite{pdg2000}, and the branching
fractions are taken separately for each form-factor as listed in
Fig.~\ref{fig:brrho}. The combined central value is determined by
taking the weighted mean of all form-factor results. The statistical
and systematic errors of the final combined result are determined by
taking the mean of the relative errors of each individual result. The
theoretical error is taken to be one half of the full spread of all
fit results (including theoretical errors). The results for each
form-factor and the combined result is shown in Fig.~\ref{fig:vub}.

\begin{figure}[htb]
\begin{center}
\epsfig{file=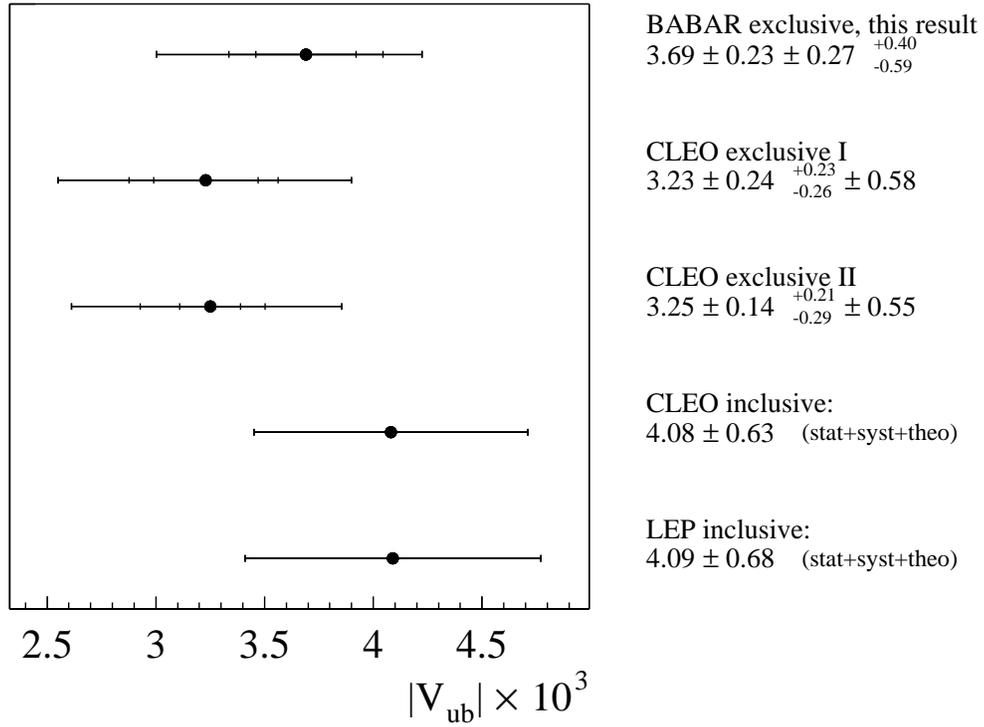,width=14cm}
\caption[Comparison with results from other experiments.]{Comparison with 
results from other experiments. The CLEO exclusive I result \cite{dlange} is obtained from 
an analysis very similar to the analysis presented here, their exclusive II result \cite{pilnu}
is an average of the exclusive I result and a separate CLEO result.}
\label{fig:vubcomp}
\end{center}
\end{figure}

A comparison of our preliminary result with inclusive and exclusive measurements
from CLEO and the inclusive measurement from LEP is shown in
Fig.~\ref{fig:vubcomp}. Two exclusive results from CLEO are quoted.
The first result is obtained from an analysis very similar to the
analysis presented here~\cite{dlange}, the second result is an average
of their first result and a separate analysis~\cite{dlange,pilnu}.
Our result is compatible with all other measurements within errors and
lies between the CLEO and LEP results.

\clearpage

\section{Acknowledgements}

We are grateful for the 
extraordinary contributions of our \pep2\ colleagues in
achieving the excellent luminosity and machine conditions
that have made this work possible.
The success of this project also relies critically on the 
expertise and dedication of the computing organizations that 
support \babar.
The collaborating institutions wish to thank 
SLAC for its support and the kind hospitality extended to them. 
This work is supported by the
US Department of Energy
and National Science Foundation, the
Natural Sciences and Engineering Research Council (Canada),
Institute of High Energy Physics (China), the
Commissariat \`a l'Energie Atomique and
Institut National de Physique Nucl\'eaire et de Physique des Particules
(France), the
Bundesministerium f\"ur Bildung und Forschung and
Deutsche Forschungsgemeinschaft
(Germany), the
Istituto Nazionale di Fisica Nucleare (Italy),
the Research Council of Norway, the
Ministry of Science and Technology of the Russian Federation, and the
Particle Physics and Astronomy Research Council (United Kingdom). 
Individuals have received support from 
the A. P. Sloan Foundation, 
the Research Corporation,
and the Alexander von Humboldt Foundation.


\begin{thebibliography}{99}

\bibitem{isgw2} D. Scora and N. Isgur, CEBAF Preprint No. CEBAF-TH-94-14.
\bibitem{beyer98} M. Beyer and D. Melikhov, \pl {\bf B436}, 344 (1998).
\bibitem{ukqcd} L. Del Debbio {\sl et al.},  \pl {\bf B 416}, 392 (1998).
\bibitem{lcsr} P. Ball and V.M. Braun, {\sl Phys. Rev.} {\bf D58}, 094016 (1998).
\bibitem{ligeti} Z. Ligeti and M.B. Wise, {\sl Phys. Rev.} {\bf D53}, 4937 (1996).
\bibitem{nim} BABAR Collaboration, B. Aubert {\sl et al.}, {\sl Nucl. Instr. and Methods} {\bf A479}, 1 (2002).
\bibitem{pepii}PEP II, SLAC-418, LBL-5379 (1993).
\bibitem{dlange} CLEO Collaboration, B.H. Behrens {\sl et al.}, {\sl Phys. Rev.} {\bf D61}, 052001 (2000).
\bibitem{foxw} G.C.~Fox and S.~Wolfram,  {\sl Nucl. Phys.} {\bf B149}, 413 (1979).
\bibitem{LAT} A. Drescher {\sl et al.}, {\sl Nucl. Instr. and Meth.} {\bf A237}, 464 (1985).
\bibitem{pdg2000} Particle Data Group, D.E. Groom {\sl et al.}, \epjc {\bf 15}, 1 (2000).
\bibitem{HQET} I.I. Bigi, M. Shifman, and N.G. Uraltsev, 
 {\sl Annu. Rec. Nucl. Part. Sci.} {\bf 47}, 591 (1997).
\bibitem{goity}J.L. Goity and W. Roberts, \pr {\bf D51}, 3459 (1995).
\bibitem{Xu}F. Fazio and M. Neubert, 
  {\em $B \rightarrow X_u \ell \nu$ decay distributions to order $\alpha_s$}, hep-ph/9905351v2.
\bibitem{iso} J.L. Diaz-Cruz, G. Lopez Castro and J.H. Munoz, \pr {\bf D54}, 2388 (1996).
\bibitem{dlange2} D.J. Lange, {\em Rare exclusive semileptonic decays at CLEO}, Ph.D. Thesis, University of California, Santa Barbara (1999).
\bibitem{barlow} R.J. Barlow and C. Beeston, {\sl Comp. Phys. Comm.} {\bf 77}, 219 (1993).
\bibitem{pilnu} J.P. Alexander {\sl et al.}, \prl {\bf 77}, 5000 (1996).

\end{thebibliography}
\end{document}